\documentclass{article}
\usepackage[utf8]{inputenc}

\title{Internal review on the state of loss of vacuum research on helium cryogenic systems}
\author{Dalesandro, Andrew A.}
\date{19 March 2023}

\begin{document}

\maketitle

\begin{abstract}
\noindent It is commonplace in helium cryogenic systems to utilize vacuum insulation to mitigate convective heat transfer to the low temperature fluids. While the insulating vacuum meaningfully improves cryogenic system thermal performance, failure of the insulating vacuum results in rapid heat transfer to the cryogenic helium often resulting in loss of fluids from venting, potential over-pressure of the system, and risks damage to vessels and equipment. The nature of the loss of vacuum problem is highly transient and difficult to predict uniformly across applications, although several efforts have incurred in recent decades to evaluate the heat transfer mechanism and develop useful models.
\end{abstract}
\section{Introduction}
The loss of insulating vacuum on helium cryogenic systems is often considered a worst-case failure scenario due a variety of confluent factors and risks. Risks include loss of life, personnel injury, irreparable damage to process equipment or vacuum vessel cryostat, inoperability of equipment, venting loss of helium storage, and reduced performance of equipment or insulation.
\subsection{Vacuum differential to ambient}
An external leak or failure would result in loss of insulating vacuum by air due to a differential pressure between the insulating vacuum and ambient air of ~ 1 atm; resulting in a high flow of air into the vacuum space. The sudden rush of air entering the cryostat may damage the  vacuum vessel or multi-layer insulation, reducing thermal performance for future operations.
\subsection{Vacuum differential to process}
It is also possible that the insulating vacuum can be spoiled from the helium process itself which may result in orders of magnitude higher pressure differential (helium process dependent) which could significantly increase the flow rate into the vacuum vessel. In addition to spoiling the vacuum, the leaking helium entering the cryostat may also excessively cool the cryostat vacuum vessel, which is typically not rated for low temperatures.
\subsection{Temperature differential to process}
Since cryogenic helium systems are largely fully welded systems inside the cryostat, loss of vacuum by air is typically the more credible failure. The large temperature differential between warm air (~300 K) and helium (e.g. liquid helium ~ 4.2 K) causes the air influx to rapidly condense and desublimate on cold helium surfaces.
\subsection{Low surface heat capacity}
Helium cryogenic systems are primarily fabricated from austenitic stainless steels due to good mechanical properties at low temperatures, including relatively high ductility. Heat capacity of austenitic stainless steels is low at low temperatures, such that the material surfaces will rapidly warm to the temperature of the condensing air on the cold surfaces, increasing the rate of heat transfer to helium.  
\subsection{Enthalpy of process helium}
The enthalpy change of the air from ambient to desublimation is > 20x the latent heat of helium at its normal boiling point. This results in a rapid  vaporization and expansion of two phase helium systems, resulting in pressurization of the system.
\subsection{Expansion ratio}
The expansion of helium from its normal boiling point to standard temperature and pressure is ~ 700x. The expansion of helium may lead to over-pressure of the system, exercising of pressure safety relief valves, and venting of helium into  ambient spaces. Due to the high volumetric expansion the venting helium may displace large volumes of air, inert the ambient environment, and create an oxygen deficient environment and potential asphyxiation. 

\section{Approaches}
Several attempts have been made to analyze and model the loss of vacuum effects on cryogenic helium systems, most commonly liquid helium systems at normal boiling point ~ 4.2 K and superfluid helium ~ 2 K. 
\subsection{Bulk vessels}
The simplest representation of the loss of vacuum problem is the bulk vessel approach in which liquid helium is stored within a vacuum insulating vessel and venting of the vacuum with air allows for the study of heat transfer and propagative effects in bulk helium. 

Perhaps the most well known and cited cryogenic helium loss of vacuum study is by Lehman and Zahn, who tested the venting of air into the vacuum space of a 4.2 K liquid helium vacuum insulated vessel, with and without multi-layer insulation (MLI) on the inner vessel~\cite{ref1}. Mass flow of air into the vacuum space and venting of helium from the inner vessel was measured and used to calculate the overall heat transfer from air to the helium. The results concluded a peak heat flux without MLI up to 3.8 W/cm$^2$ while the MLI wrapped vessel limited peak heat flux to 0.6 W/cm$^2$, thus indicating that use of MLI on liquid helium cooled surfaces can significantly mitigate loss of vacuum impacts.

A similar experiment was conducted by Harrison et al by venting air to the insulating vacuum space on a bulk helium vessel both with and without composite insulation, but instead with superfluid helium at 1.8 K~\cite{ref2}. Observed peak heat flux without and with insulation was 3.1 W/cm$^2$ and 0.44 W/cm$^2$, respectively, indicating again that insulating the internal vessel has significant mitigative effects on heat transfer to liquid helium due to loss of vacuum, and agreeing rather well with the Lehman and Zahn results.

Another set of researchers at KIT developed an analytical model for sudden loss of insulating vacuum by air~\cite{ref3} and designed a liquid helium pressurized cryostat test facility named PICARD to help validate the model~\cite{ref4} testing loss of vacuum heat transfer of a normal boiling liquid helium volume of 100 liters, operating pressure of 16 barg, and a helium mass flow vent rates $\le$ 4 kg/s~\cite{ref5},\cite{ref6}. Results from PICARD concluded that relief valves sized for 4 W/cm$^2$ tend to be oversized, resulting in chatter and possible overpressure due to reduced relief capacity. Further results also concluded that while peak heat flux can exceed 5 W/cm$^2$, the nominal heat flux decays quickly to $<$ 2 W/cm$^2$ due to reduced heat transfer rates from gas desublimation on the cold surfaces~\cite{ref7},\cite{ref8}.

\subsection{Representative geometry}
Due to the complex dynamics at work during a loss of insulating vacuum on a liquid helium cryogenic system and fast  temperature, pressure, and heat transfer transients, attempts have been made to use more representative geometry to test loss of vacuum effects for specific applications rather than relying on results from bulk vessel tests. 

Superconducting radio frequency (SRF) cavities for particular acceleration are one of the most prominent applications for large scale superfluid helium usage and over-pressure safety control tends to be strongly influenced by the very high geometric aspect ratios of these devices (length $>>$ diameter). Experiments on scaled SRF cavities in which the beamline vacuum is vented to air have concluded that this high aspect ratio reduces the rate of heat transfer to the superfluid by cryo-pumping the inflowing air on the cold pipe inner surface, effectively slowing the speed of air in the beam-line to $\sim$10 m/s, well below the air's speed of sound, even at low temperatures~\cite{ref9}.

\subsection{Simplified geometry}
Instead of designing and developing experiment test geometry comparable to production cryogenic system geometry to mitigate uncertainty in the transient effects, another approach is to reduce the number of variables and study the loss of insulating vacuum on simplified geometry.

One of the most systematic approaches to the study of loss of insulating vacuum in normal boiling point liquid helium were performed by Bosque~\cite{ref10} and Dhuley~\cite{ref11} in which the rate of air deposition during a loss of vacuum event is measured across a liquid wetted copper disc. These tests concluded that the air condensation rate quickly peaks as the vacuum is vented, but also decays quickly as the total deposition of air steadily increases, while the rate of heat transfer to the LHe bath is limited by film boiling. Dhuley and Van Sciver~\cite{ref12},\cite{ref13} continued these experiments with an evacuated vertical tube immersed in LHe, and instrumented longitudinally with pressure and temperature transmitters to accurately measure the air deposition propagation rate, and calculated an effective heat flux $\sim$2.3 W/cm$^2$ based on the pool boiling heat transfer curve for 4.2 K saturated LHe~\cite{ref14}. The authors followed up their experimental results with development of a numerical model based on parametric data across multiple vacuum venting air flow rates to predict and explain that the slow air propagation and deposition speed is a function of the reduced heat transfer rate by desublimation of air on the helium cooled tube surface and exponential decrease in air temperature along the length of the channel~\cite{ref15},\cite{ref16}. The model predicts that given a high aspect ratio (e.g. in an evacuated tube or a in a SRF beamline) there is a finite length in which the air ingress venting the evacuated space will freeze out entirely with only minimal thermal impact to the rest of the cold tube. Further studies analyzing the dynamic effects of the air deposition heat transfer concluded that while the peak heat flux from air can exceed 10 W/cm$^2$, the peak heat flux is actually limited by film boiling in LHe to $\le$ 2.5 W/cm$^2$~\cite{ref17}-\cite{ref20}.

Continuing the work of Dhuley et al, Garceau et al performed comparable experiments with an extended length of evacuated tubes and performed tests with superfluid helium~\cite{ref21}-\cite{ref24}. Bao et al developed a representative theoretical model of Garceau's experimental results~\cite{ref25}.

\subsection{Field tests}
While several options exist to model and simulate loss of insulating vacuum in liquid helium systems, the most representative way to understand the transient effects and heat transfer rate for a specific system design is to perform a field test on a full-scale prototype of an as-built system.

For SRF cryo-module cavity applications, two such sets of loss of vacuum field experiments have been performed at full-scale. Tests at CEBAF at Thomas Jefferson National Laboratory in 1994 studied loss of beamline vacuum on a quarter SRF cryomodule at both 2 K and 4 K LHe and measured peak heat flux of 3.5 W/cm$^2$ and 2.8 W/cm$^2$, respectively, with a maximum sustained heat flux of 2 W/cm$^2$~\cite{ref26}. Similarly, tests at XFEL at Deutsches Elektronen-Synchrotron (DESY) in 2008 measured the heat flux to helium from loss of insulating vacuum and loss of beam vacuum at 2 K and at 4.4 K with a full-scale, complete XFEL SRF cryomodule~\cite{ref27}. The measured heat flux due to loss of insulating vacuum at  2 K and 4.4 K was $\le$ 0.4 W/cm$^2$ and $\le$ 0.65 W/cm$^2$, respectively while the loss of beam vacuum heat flux was $\le$ 2.3 W/cm$^2$. The experiments also concluded substantial delay in air pressure propagation during venting the beam vacuum of $\sim$3 m/s based on a $>$ 4 seconds delayed in pressure pickup between pressure transmitters at the near and far end of $\sim$12 m SRF cryomodule beam line.

\section{Codes and standards}

Loss of vacuum influences on system design typically manifests in overpressure protection. Design codes and standards provide methodology for sizing relief devices and vent lines to prevent over-pressure of piping and pressure vessels.
Notable codes include Compressed Gas Association S1-3~\cite{ref28} and International Standards Organization ISO-EN-21013~\cite{ref29}. ISO-EN-21013 provides guidance to estimate heat load to LHe for MLI and uninsulated vessels. CGA S1-3 specifies calculating heat load based on the surface area of the helium containing vessel, the thickness of thermal insulation, and thermal conductivity of the insulation, which for vacuum insulated vessels is not always appropriate. For dynamic thermofluid properties of helium, reference is made to National Bureau of Standards Technology of Liquid Helium~\cite{ref30}.

\section{Conclusion}

Loss of insulating vacuum, and loss of beam vacuum for SRF cavity applications, is a reasonably plausible failure mode for liquid helium systems which often is considered a worst-case failure. There have been significant efforts to study and understand the dynamic heat flux and transient effects of air deposition and propagative affects, especially for high aspect ratios geometries. Generally for LHe systems, loss of insulating vacuum can be predicted between $\sim$2-4 W/cm$^2$ for uninsulated vessels and $\sim$0.6 W/cm$^2$ for MLI insulated vessels, while loss of beam vacuum failures can be predicted between $\sim$1.4-2.3 W/cm$^2$ since beam lines are typically uninsulated.


\begin{thebibliography}{40}
\bibitem{ref1}
W. Lehmann and G. Zahn, “Safety Aspects for the LHe Cryostats and LHe Transport Containers,” Proc. of the ICEC 7, London, IPC Science and Technology Press, pp. 569-579, 1978. 
\bibitem{ref2}
S. M. Harrison, “Loss of vacuum experiments on a superfluid helium vessel,” IEEE Transactions on Applied Superconductivity 12(1), 1343 – 1346, 2002. 
\bibitem{ref3}
C. Heidt, S. Grohmann, and M. Süßer, “Modeling the pressure increase in liquid helium cryostats after failure of the insulating vacuum,” AIP Conference Proceedings 1573 (1), 1574-1580, 2014. 
\bibitem{ref4}
C. Heidt, H. Schön, M. Stamm, and S. Grohmann, “Commissioning of the cryogenic safety test facility PICARD,” IOP Conference Series: Materials Science and Engineering 101 (1), 012161, 2015. 
\bibitem{ref5}
C. Heidt, A. Henriques, M. Stamm, and S. Grohmann, “First experimental data of the cryogenic safety test facility PICARD,” IOP Conference Series: Materials Science and Engineering 171, 012044, 2017. 
\bibitem{ref6}
C. Heidt, “Experimental Investigation and Modelling of Incidents in Liquid Helium Cryostats,” PhD Dissertation, Karlsruhe Institute of Technology, 2018. 
\bibitem{ref7}
C. Weber, A. Henriques, C. Zoller, and S. Grohmann, “Safety studies on vacuum insulated liquid helium cryostats,” IOP Conference Series: Materials Science and Engineering 278, 012169, 2017. 
\bibitem{ref8}
C. Weber, A. Henriques, and S. Grohmann, “Study on the heat transfer of helium cryostats following loss of insulating vacuum,” IOP Conference Series: Materials Science and Engineering 502, 012170, 2019. 
\bibitem{ref9}
Andrew A. Dalesandro, Ram C. Dhuley, Jay C. Theilacker, and Steven W. Van Sciver, “Results from sudden catastrophic loss of vacuum on scaled superconducting radio frequency cryomodule”, AIP Conference Proceedings 1573, 1822-1829, 2014.
\bibitem{ref10}
E. S. Bosque, R. C. Dhuley, and S. W. Van Sciver, “Transient heat transfer in Helium II due to sudden vacuum break”, AIP Conference Proceedings 1573, 260-267, 2014. 
\bibitem{ref11}
R. C. Dhuley, E. S. Bosque, and S. W. Van Sciver, “Cryodeposition of nitrogen gas on a surface cooled by helium II,” AIP Conference Proceedings 1573, 626-632, 2014. 
\bibitem{ref12}
Ram C. Dhuley and Steven W. Van Sciver, “Sudden vacuum loss in long liquid helium cooled tubes", IEEE Transactions on Applied Superconductivity 25(3), 9000305, 2015. 
\bibitem{ref13}
R. C. Dhuley and S. W. Van Sciver, “Epoxy encapsulation of a CernoxTM SD thermometer for measuring the temperature of surfaces in liquid helium”, Cryogenics 77, 49-52, 2016. \bibitem{ref14}
W. G. Steward, “Transient helium heat transfer: Phase I– Static coolant,” International Journal of Heat and Mass Transfer 21, 863–874, 1978. 
\bibitem{ref15}
R. C. Dhuley and S. W. Van Sciver, “Propagation of nitrogen gas in a liquid helium cooled vacuum tube following sudden vacuum loss- Part I: Experimental investigations and analytical modeling”, International Journal of Heat and Mass Transfer 96, 573-581, 2016. 
\bibitem{ref16}
R. C. Dhuley and S. W. Van Sciver, “Propagation of nitrogen gas in a liquid helium cooled vacuum tube following sudden vacuum loss- Part II: Analysis of propagation speed”, International Journal of Heat and Mass Transfer 98, 728-737, 2016. 
\bibitem{ref17}
R. C. Dhuley and S. W. Van Sciver, “Heat transfer in a liquid helium cooled vacuum tube following sudden vacuum loss”, IOP Conference Series: Material Science and Engineering 101, 012006, 2015. 
\bibitem{ref18}
R. C. Dhuley and S. W. Van Sciver, “Nitrogen gas propagation in a liquid helium cooled vacuum tube following a sudden vacuum loss,” IOP Conference Series: Material Science and Engineering 171, 012084, 2015. 
\bibitem{ref19}
R. C. Dhuley, S. Posen, O. Prokofiev, M. I. Geelhoed, and J. C. T. Thangaraj, “First demonstration of a cryocooler conduction-cooled superconducting radiofrequency cavity operating at practical cw accelerating gradients”, Superconductor Science and Technology 33, 06LT01, 2020.  
\bibitem{ref20}
R. C. Dhuley, R. Kostin, O. Prokofiev, M. I. Geelhoed, T. H. Nicol, S. Posen, J. C. T. Thangaraj, T. K. Kroc, and R. D. Kephart, “Thermal link design for conduction cooling of a superconducting radio frequency cavity using cryocoolers”, IEEE Transactions on Applied Superconductivity 29(5), 0500205, 2019. 
\bibitem{ref21}
N. Garceau, S. Bao, and W. Guo, “Effect of mass flow rate on gas propagation after vacuum break in a liquid helium cooled tube,” IOP Conference Series: Material Science and Engineering 755, 012112, 2015. 
\bibitem{ref22}
N. Garceau, S. Bao, and W. Guo, “Heat and mass transfer during a sudden loss of vacuum in a liquid helium cooled tube – Part I: Interpretation of experimental observations,” International Journal of Heat and Mass Transfer 129, 1144-1150, 2019. 
\bibitem{ref23}
N. Garceau, S. Bao, W. Guo, and S. W. Van Sciver, “The design and testing of a liquid helium cooled tube system for simulating sudden vacuum loss in particle accelerators,” Cryogenics 100, 92-96, 2019. 
\bibitem{ref24}
N. Garceau, S. Bao, W. Guo, “Heat and mass transfer during a sudden loss of vacuum in a liquid helium cooled tube - Part III: Heat deposition in He II,” International Journal of Heat and Mass Transfer 181, 121885, 2021. 
\bibitem{ref25}
S. Bao, N. Garceau, and W. Guo, “Heat and mass transfer during a sudden loss of vacuum in a liquid helium cooled tube – Part II: Theoretical modeling,” International Journal of Heat and Mass Transfer 146, 118883, 2020. 
\bibitem{ref26}
M. Wiseman, K. Crawford, M. Drury, K. Jordan, J. Preble, Q. Saulter, and W. Schneider, “Loss of cavity vacuum experiment at CEBAF,” Advances in cryogenic engineering, Volume 39(A), 997-1003, 1994. 
\bibitem{ref27}
T. Boekman et al., “Experimental tests of fault conditions during the cryogenic operation of a XFEL prototype cryomodule,” Proc. ICEC 22- ICMC 2008, 22:723–728, 2009.
\bibitem{ref28}
Compressed Gas Association, “S.1: Pressure Relief Device Standards” 
\bibitem{ref29}
Cryogenic vessels — Pressure-relief accessories for cryogenic service — Part 3: Sizing and capacity determination. 
\bibitem{ref30}
B. W. Birmingham and R. H. Kropschot, “Technology of liquid helium”, 1968. 
\end{thebibliography}
\end{document}